\documentclass[preprint,aps,tightenlines,superscriptaddress]{revtex4}
\usepackage{graphicx}
\usepackage{bm}
\usepackage{epsfig}
\newif\ifpdf
\ifx\pdfoutput\undefined
\pdffalse 
\else
\pdfoutput=1 
\pdftrue
\fi

\newcommand{\bea}{\begin{eqnarray}}
\newcommand{\eea}{\end{eqnarray}}
\newcommand{\beq}{\begin{equation}}
\newcommand{\eeq}{\end{equation}}
\newcommand{\bay}{\begin{array}}
\newcommand{\eay}{\end{array}}

\begin{document}
\ifpdf
\DeclareGraphicsExtensions{.pdf, .jpg}
\else
\DeclareGraphicsExtensions{.eps, .jpg}
\fi
\vspace{1.5cm}
\preprint{ \vbox{\hbox{UCSD/PTH 02-20} \hbox{hep-ph/0209211}   }}
\vspace{2.0cm}

\title{Subleading corrections to the $|V_{ub}|$ determination from 
exclusive $B$ decays} 
\author{Benjam\'\i{}n Grinstein and Dan Pirjol\footnote{Present address:
Department of Physics and Astronomy, The Johns Hopkins University, 3400 North
Charles Street, Baltimore, MD 21218}}
\affiliation{Department of Physics, UCSD, 9500 Gilman Drive, La Jolla, CA 92093}

\date{\today\\ \vspace{2cm} }

\begin{abstract}
\vspace{1.0cm}
\setlength\baselineskip{18pt}

It has been proposed to determine the CKM matrix element $|V_{ub}|$
in a model-independent way from a combination of rare and semileptonic 
$B$ and $D$ decays near the zero recoil point. An essential ingredient
in such a determination is a
heavy quark symmetry relation connecting the form-factors appearing in
$B\to K^*e^+ e^-$ to semileptonic form factors relevant for
$B\to \rho e\nu$. We estimate the leading corrections to this
symmetry relation, of order $\alpha_s(m_b)$ and $\Lambda/m_b$, pointing
out that they can be as large as 20\%, depending on the value of the
matrix element of a dimension-4 operator. Dimensional analysis estimates
of this matrix element give a corresponding uncertainty in $|V_{ub}|$ of the
order of a few percent.
\end{abstract}

\maketitle


The Cabibbo-Kobayashi-Maskawa matrix element $V_{ub}$ is an important
ingredient of the flavor structure of the Standard Model. Its magnitude
determines one of the sides of the unitarity triangle.
Although at present it is one of most poorly known CKM parameters, several
promising methods have been proposed to extract it from experimental data
(see \cite{vub} for a recent review).

We focus in this paper on a particular model-independent
method for determining $|V_{ub}|$
from exclusive rare and semileptonic $B$ decays proposed in \cite{SaYa,LiWi,LSW}.
There are two basic ingredients going into such a determination:
\begin{enumerate}
\item Heavy quark symmetry relations between heavy-light form factors \cite{IW,BuDo}
connect the rate for $\bar B\to K^* e^+ e^-$ at the zero recoil point
$q^2_{\rm max}$ to the rate for $\bar B\to \rho e\nu$, up to $\Lambda/m_b$
and $SU(3)$ breaking corrections.

\bea\label{1}
\frac{\mbox{d}\Gamma(\bar B\to \rho e\nu)/\mbox{d}q^2}
{\mbox{d}\Gamma(\bar B\to K^* \ell^+ \ell^-)/\mbox{d}q^2}
= 
\frac{|V_{ub}|^2}{|V_{tb} V_{ts}^*|^2} \cdot \frac{8\pi^2}{\alpha^2}\cdot
\frac{1}{|C_9|^2 + |C_{10}|^2}
\frac{|f^{B\to\rho}(y)|^2}{|f^{B\to K^*}(y)|^2}
\cdot \frac{1}{1 + \Delta(y)}
\eea
In this relation the ratio of decay rates is taken at a common value of 
the parameter $y$ defined by $v\cdot p_V = m_V y$. The correction $\Delta(y)$ 
parametrizes the contribution of the magnetic penguin operator $Q_7$
near $y=1$.

\item The SU(3) breaking corrections introduced in step 1 can be eliminated 
with the
help of semileptonic decay $D\to (\rho, K^*) e\nu$ data, by using an approximate
equality to unity of the double ratio \cite{Grin}
\bea\label{2}
R(y) \equiv \frac{|f^{B\to\rho}(y)/f^{B\to K^*}(y)|}{|f^{D\to\rho}(y)/f^{D\to K^*}(y)|}
= 1 + {\cal O}\left(\frac{m_s}{\Lambda_\chi}(\frac{\Lambda}{m_c} - \frac{\Lambda}{m_b})
\right)\,.
\eea

\end{enumerate}

First evidence for the mode $B\to K^{*}\ell^+\ell^-$ has been recently reported
by the BaBar collaboration, and an upper limit was given by BELLE \cite{Babar,Belle}
\bea
{\cal B}(B\to K^* \ell^+ \ell^-) &=& (1.68^{+0.68}_{-0.58}\pm 0.28) 
\times 10^{-6}\quad \mbox{(BaBar)}\\
&<& 14 \times 10^{-7} \mbox{ (90\% CL)}\qquad \,\mbox{  (BELLE)}\,.\nonumber
\eea
This suggests that a determination of $|V_{ub}|$
using these decays might become feasible in a not too distant future.

There are several sources of theoretical uncertainties connected with 
such an approach. The dominant theoretical uncertainty is connected with long-distance
contributions coming from four-quark operators in the weak Hamiltonian 
$Q_1 = (\bar s_\alpha
\gamma_\mu P_L b_\beta)(\bar c_\beta \gamma^\mu P_L c_\alpha)$,
$Q_2 = (\bar s\gamma_\mu P_L b)(\bar c\gamma_\mu P_L c)$ \cite{Heff}. 
Their effect can be absorbed
into a redefinition of the Wilson coefficient $C_9$ \cite{Heff}
but this modification is generally process-dependent. It can be computed
reliably in perturbation theory as long as the $e^- e^+$ invariant mass
is sufficiently low. Such computations have been performed for inclusive 
$B\to X_se^+ e^-$
\cite{Heff,BuMu} and exclusive $B\to K^* e^+e^-$ decays \cite{BeFeSe,pQCD}. At small
recoil, as considered in this Letter, this method is not applicable and one
has to resort instead to a phenomenological parameterization of these effects
in terms of sums over $J/\psi$ resonances \cite{VMD}. While this is not a controlled
approximation, the validity of such an expansion can be tested by measuring
other observables such as $q^2$ spectra and/or angular asymmetries \cite{LSW,pQCD}.

A better understood source of uncertainty is
introduced in step 2 as corrections to the double ratio $R(y)$. From 
power counting, such corrections are of order $\frac{m_s}{\Lambda_\chi}\left(
\frac{\Lambda}{m_c} - \frac{\Lambda}{m_b}\right)\simeq 7\%$. A more precise
estimate has been performed in \cite{LSW} using the chiral perturbation theory 
for vector mesons \cite{chpthv}. The corresponding deviation of the ratio $R(1)$
from unity was found to be small, under 1\%.

Finally, another source of corrections is introduced in the first step of the method. 
As shown in \cite{SaYa}, the contribution of the electromagnetic penguin 
operator $Q_7$ to the $B\to K^* e^+ e^-$ decay rate (parameterized by 
$\Delta(y)$) can be computed in a model-independent way at the zero recoil point 
$y=1$ using heavy quark symmetry. The corrections to this prediction are of order
$O(\Lambda_{QCD}/m_b)$ and were thus thought to be negligible.

Recently, the structure of the leading corrections to the heavy quark symmetry 
relations for heavy-light decays \cite{IW,BuDo} has been studied in
\cite{GrPi}. They come from dimension 4 operators in the matching of the
weak current onto heavy quark effective theory (HQET) operators, and
from hard gluon effects appearing as Wilson coefficients in the HQET.
We are therefore now in a position to study the leading corrections
in step 1 of the $|V_{ub}|$ determination. 
We find that there are calculable corrections to the heavy quark
symmetry relations of the order of 10\%, plus an additional correction
which depends on an unknown form factor of a dimension-4 operator.
Estimates of this form factor from dimensional analysis result in an
uncertainty on $|V_{ub}|$ of the order of 3\%.

The amplitude for $\bar B\to K^* e^+ e^-$  receives in general contributions
from all operators in the weak radiative $b\to s \ell^+\ell^-$ effective Hamiltonian
\cite{Heff}. The result can be expressed in terms of the matrix elements of the
three operators
\bea
Q_7 &=& \frac{e^2 m_b}{16\pi^2}  (\bar s\sigma_{\mu\nu} P_R b) F^{\mu\nu}\\
Q_9 &=& \frac{e^2}{16\pi^2} (\bar s\gamma_\mu P_L b) (\bar e\gamma^\mu e)\\
Q_{10} &=& \frac{e^2}{16\pi^2} (\bar s\gamma_\mu P_L b) 
(\bar e\gamma^\mu\gamma_5 e)\,,
\eea
where the effects of the remaining operators $Q_{1-6}$ are understood to be
absorbed into redefinitions of the Wilson coefficients $C^{\rm eff}_{7,9,10}$.

The amplitude for the decays considered here requires the knowledge of
the form-factors of the vector, axial and tensor currents
$\bar s\Gamma b$. We define them as \cite{MaWi,IW} (with the convention
$\varepsilon^{0123} = 1$)
\bea\label{VA}
\langle V(p',\epsilon)|\bar q\gamma_\mu b|\bar B(p)\rangle &=&
ig(q^2) \varepsilon_{\mu\nu\lambda\sigma} \epsilon^*_\nu
(p+p')_\lambda (p-p')_\sigma\\
\label{4}
\langle V(p',\epsilon)|\bar q\gamma_\mu\gamma_5 b| \bar B(p)\rangle &=&
f(q^2) \epsilon^*_\mu 
+ \, a_+(q^2)(\epsilon^*\cdot p)(p+p')_\mu\\
 & &\qquad\quad\,\,\,\, +\,
a_-(q^2)(\epsilon^*\cdot p)(p-p')_\mu\,.\nonumber
\eea
and 
\bea\label{T}
\langle V(p',\epsilon)|\bar q\sigma_{\mu\nu} b|\bar B(p)\rangle &=&
g_+(q^2) \varepsilon_{\mu\nu\lambda\sigma} \epsilon^*_\lambda
(p+p')_\sigma + 
g_-(q^2) \varepsilon_{\mu\nu\lambda\sigma} 
\epsilon^*_\lambda (p-p')_\sigma\\
 & +&
h(q^2) \varepsilon_{\mu\nu\lambda\sigma} (p+p')_\lambda
(p-p')_\sigma(\epsilon^*\cdot p)\nonumber\,.
\eea

The differential decay rate for $\bar B\to K^* e^+ e^-$ can be written
as a sum of contributions corresponding to well-defined helicities $\lambda$
of the vector meson
\bea
\frac{\mbox{d}}{\mbox{d}q^2}\Gamma(\bar B\to K^* e^+ e^-) \propto
\sum_{\lambda = \pm 1, 0} \left\{ |H_\lambda^{(V)}|^2 + |H_\lambda^{(A)}|^2
\right\}\,.
\eea
The two helicity amplitudes $H_\lambda^{(V,A)}$ correspond to the vector and
axial coupling to the leptons, respectively.
Expressed in terms of the form factors defined in (\ref{VA})--(\ref{T}), they
are given explicitly by
\bea
H_{\pm 1}^{(V)}(q^2) &=& C_7\frac{2m_b}{q^2}\left(g_+(q^2)
[\pm 2m_B |\vec q\,| + m_B^2 - m_V^2] + g_-(q^2) q^2\right)\\
&+& C_9 [\pm 2m_B g(q^2) |\vec q\,| + f(q^2)]\nonumber\\
H_0^{(V)}(q^2) &=& C_7\frac{2m_b}{q^2}\frac{\sqrt{q^2}}{2m_V}
\left(g_+(q^2) (m_B^2 + 3m_V^2 - q^2) + g_-(q^2) (m_B^2 - m_V^2 - q^2)
+ 4h(q^2) m_B^2 \vec q\,^2\right)\nonumber\\
&+& C_9 \frac{1}{m_V\sqrt{q^2}}
\left( f(q^2) \frac12 (m_B^2-m_V^2-q^2) + 2m_B^2 \vec q\,^2 a_+(q^2)\right)
\eea
and
\bea
H_{\pm 1}^{(A)}(q^2) &=& C_{10} (\pm 2m_B g(q^2) |\vec q\,| + f(q^2))\\
H_{0}^{(A)}(q^2) &=& C_{10} \frac{1}{m_V\sqrt{q^2}}
\left( f(q^2) \frac12 (m_B^2-m_V^2-q^2) + 2m_B^2 \vec q\,^2 a_+(q^2)\right)
\eea
At the zero recoil point $y=1$ (corresponding to maximal $q^2$, 
$q_{\rm max}^2 = (m_B-m_V)^2$),
the form of the helicity amplitudes simplifies drastically and
can be written only in terms of one axial form factor
$f(y=1)$ as
\bea\label{HV1}
H_{\pm 1}^{(V)}(y=1) &=& H_{0}^{(V)}(y=1) = C_9 f(1) (1 + \delta(1))\\
\label{HA1}
H_{\pm 1}^{(A)}(y=1) &=& H_{0}^{(A)}(y=1) = C_{10} f(1)
\eea
where we defined
\bea
\delta(y) \equiv \frac{2m_b}{m_B - m_V}\frac{C_7}{C_9}\cdot \frac{{\cal F}(y)}{f(y)}
\eea
with 
\bea
{\cal F}(y) \equiv g_+(y) (m_B + m_V) + g_-(y) (m_B - m_V)\,.
\eea

Inserting the expressions for the helicity amplitudes at the end-point
(\ref{HV1}), (\ref{HA1}) into the rate for $B\to K^* \ell^+\ell^-$, gives 
the following result for the correction $\Delta(y)$ appearing in (\ref{1})
\cite{SaYa,LiWi}
\bea\label{delta}
\Delta(y) = \frac{1}{|C_9|^2 + |C_{10}|^2}\left\{
\frac{4m_b}{m_B-m_V}\mbox{Re }(C_7 C_9^*) \frac{{\cal F}(y)}{f(y)} +
\frac{4m_b^2}{(m_B-m_V)^2}|C_7|^2 \left(\frac{{\cal F}(y)}{f(y)}\right)^2
\right\}\,.
\eea

The form-factor ratio ${\cal F}(1)/f(1)$ is predicted from heavy quark 
symmetry to be 1 at leading order in the heavy quark limit \cite{IW,SaYa}.
The purpose of this Letter is to estimate the leading corrections to this result,
and study their effect on the $|V_{ub}|$ determination. 

Using the results of \cite{GrPi} one finds the following prediction
from heavy quark symmetry
\bea\label{20}
{\cal F}(y) &=& \left( \kappa_5 + \frac{\bar\Lambda - m_V}{m_B}\right)
f(y) + 2m_B m_V \left(1 - \frac{1}{\kappa_1}\right) g(y) +
\frac{2}{m_B}{\cal D}_1(y)\\
&+& 2m_B m_V \left( 1 - \frac{\bar\Lambda + m_V}{m_B}\right)(y-1) g(y) +
2m_V (y-1) {\cal D}(y) + \cdots\nonumber\,,
\eea
where the ellipses denote contributions suppressed by $\Lambda^2/m_b^2$ relative
to the leading term.

The subleading form factors ${\cal D}(y)$ and ${\cal D}_1(y)$ appearing in 
(\ref{20}) are defined by matrix elements of the dimension-4 currents
\bea\label{Ddef}
\langle V(p',\varepsilon) |\bar qiD_\mu h_v|\bar B(v)\rangle &=&
{\cal D}(y) i\varepsilon_{\mu\nu\lambda\sigma} \varepsilon_\nu^* p_\lambda
p'_\sigma\\
\label{D1def}
\langle V(p',\varepsilon) |\bar qiD_\mu \gamma_5 h_v|\bar B(v)\rangle &=&
{\cal D}_1 \varepsilon^*_\mu + {\cal D}_+ (\varepsilon^*\cdot p)(p_\mu +
p'_\mu) + {\cal D}_- (\varepsilon^*\cdot p)(p_\mu -p'_\mu)\,.
\eea
The equation of motion for the heavy quark field $iv\cdot D h_v= 0$ implies a relation
among the form factors of the $\bar qiD_\mu\gamma_5 h_v$ current, such that 
only two of them are independent.

The coefficients $\kappa_1$ and $\kappa_5$ contain hard gluon corrections.
The first coefficient $\kappa_1 = -c_0(m_b)/c'_0(m_b)$ is defined as the ratio of 
two Wilson 
coefficients appearing in the matching for $J_\mu = \bar q\gamma_\mu b$ and 
$J'_\mu = \bar qi\sigma_{\mu\nu} v^\nu b$
\bea
J^{(\prime )}_{\mu} = 
c_0^{(\prime )}(\mu) \bar q\gamma_\mu h_v + 
c_1^{(\prime )}(\mu) \bar qv_\mu h_v
+ O(1/m_b)\,.
\eea
The values of the Wilson coefficients can be extracted from the next-to-leading 
computation of \cite{BrGr} and their explicit values can be found in \cite{GrPi}.  
For our estimate we only require their expressions to one-loop order, which 
give $\kappa_1 = 1 + O(\alpha^2_s(m_b))$.

The coefficient $\kappa_5$ is defined analogously in terms of the Wilson 
coefficients appearing in the matching of the
currents $J_{5\mu} = \bar q i\sigma_{\mu\nu}v^\nu \gamma_5 b$ and
$J'_{5\mu} = (g_{\mu\nu} - v_\mu v_\nu) \bar q \gamma^\nu \gamma_5 b$
\bea
J^{(\prime )}_{5\mu} = 
\tilde c_0^{(\prime )}(\mu) \bar q\gamma_\mu\gamma_5 h_v + 
\tilde c_1^{(\prime )}(\mu) \bar qv_\mu \gamma_5 h_v
+ O(1/m_b)\,.
\eea\
The explicit results for the Wilson coefficients depend on the $\gamma_5$ definition
and are given by \cite{BrGr}
\bea\label{c5match}
& &\tilde c_0(m_b) = \tilde c_1(m_b) = 1 - \frac{4\alpha_s(m_b)}{3\pi} + O(\alpha_s^2)
\quad \mbox{(NDR,HV)}\\
& &\tilde c'_0(m_b) = \tilde c'_1(m_b) = \left\{
\begin{array}{cc}
1 + O(\alpha_s^2)
& \mbox{(HV)} \\
1 - \frac{4\alpha_s(m_b)}{3\pi} + O(\alpha_s^2) & \mbox{(NDR)}
\end{array}
\right.
\eea
This gives for the coefficient $\kappa_5$ 
\bea
\kappa_5 = \frac{\tilde c_0(m_b)}{\tilde c'_0(m_b)} = 
\left\{
\begin{array}{cc}
1 + O(\alpha_s^2) & \mbox{(NDR)} \\
1 - \frac{4\alpha_s(m_b)}{3\pi} + O(\alpha_s^2) & \mbox{(HV)}
\end{array}
\right.
\eea

For completeness we give also the corrected HQET symmetry relations used in deriving 
(\ref{20}) \cite{GrPi}
\bea\label{Icorr}
& &\kappa_1 (g_+ - g_-) + 2m_B g =
-2(m_V y - \bar\Lambda) g(y) - \frac{1}{m_B} f(y) - 2 {\cal D}(y) + 
{\cal O}(m_b^{-3/2})\\
\label{IIcorr}
& &
g_+ + g_- - 2m_V y g - \kappa_5 \frac{1}{m_B}f =\\
& &\qquad\quad -2 \frac{m_V}{m_B}(\bar\Lambda y - m_V) g(y) + \frac{\bar\Lambda}{m_B^2}
f + 
\frac{2}{m_B^2}(
m_V m_B y{\cal D}(y) + {\cal D}_1(y)) + {\cal O}(m_b^{-5/2})\nonumber
\eea

From (\ref{20}) one finds for the ratio of form factors appearing in $\delta(1)$ 
at the zero recoil point $y=1$
\bea\label{F1f1}
\frac{{\cal F}(1)}{f(1)} &=& \kappa_5  + \frac{\bar\Lambda - m_V}{m_B} +
2m_B m_V\left(1 - \frac{1}{\kappa_1}\right)\frac{g(1)}{f(1)} +
\frac{2}{m_B}\cdot \frac{{\cal D}_1(1)}{f(1)}\\
&=& \left\{
\begin{array}{cc}
1 - 0.1 + \frac{2}{m_B}\cdot \frac{{\cal D}_1(1)}{f(1)} & \mbox{(NDR)} \\
1 - 0.09 - 0.1 + \frac{2}{m_B}\cdot \frac{{\cal D}_1(1)}{f(1)} & \mbox{(HV)}
\end{array}
\right. \nonumber
\eea
This is the main result of this Letter. We turn now to a discussion of the
numerical impact of these corrections.

In the NDR scheme, the radiative corrections from $\kappa_{1,5}$
start at two-loop order and thus are negligibly small (the numerical 
value for $\kappa_5^{(HV)}$ uses $\alpha_s(m_b) = 0.22$). The subleading
$1/m_b$ correction from the second term is a negative $\sim 10\%$ contribution,
and is kinematically enhanced by the large ratio 
$(m_{K^*}-\bar\Lambda)/m_B \simeq 0.10$, where we took $\bar\Lambda = 350$ MeV.
The last term depends on the subleading
form factor ${\cal D}_1(1)$, which has not been calculated yet. From dimensional
analysis its contribution is expected to be of order $\Lambda/m_b \simeq 0.10$.
Therefore, if this term turns out to be negative, the overall correction 
to the ratio (\ref{F1f1}) in the NDR scheme could be as large as $-20\%$. 

We give in Table I
the values of the coefficient $\Delta(1)$ corresponding to a generic range
for the form factor ratio (\ref{F1f1}), using two sets of Wilson coefficients 
$C_{7,9,10}(m_b)$.
The first set corresponds to the next-to-leading log (NLL) approximation
$C_7^{LL} = -0.314$, $C_9^{NLL}=4.154$, $C_{10}^{NLL} = -4.261$. 
These values correspond to the NDR scheme, and were obtained at $\mu = 4.6$ GeV,
with $\bar m_b(\bar m_b) = 4.4$ GeV, $\Lambda_{QCD}^{(n_f=5)} = 220$ MeV following
\cite{BeFeSe}.
The perturbative correction from $\kappa_5$ multiplies $C_7$,  thus at this 
order consistency requires that $\kappa_5$ not be included.
The one-loop matrix elements of $Q_{1-6}$ are included by
absorbing them into a scheme-independent effective Wilson coefficient
$C_9^{\rm eff}$ defined as in \cite{Heff,BuMu}. At the zero recoil point
this is given by $C_9^{\rm eff}(1) = C_9(m_b)\tilde \eta(1) + 0.208 + i 0.363$,
with $\tilde \eta(1) = 1 - 0.61\alpha_s(m_b)/\pi = 0.95$, where we used $m_c = 1.4$ GeV.
The corresponding numerical results for $\Delta(1)$ are shown in the first line 
of Table I.

A second set of Wilson coefficients corresponds to the NNLL approximation
and takes $C_7^{NLL} = -0.308$, $C_9^{NNLL}=4.214$, $C_{10}^{NNLL} = -4.312$
\cite{BeFeSe,NNLL}. These partial results corresponding to the NDR scheme do not
contain the (as yet unknown) three-loop mixing into $Q_9$. The associated
uncertainty in $C_9^{NNLL}$ was estimated in \cite{BeFeSe} and found to be 
$\sim \pm 0.1$. 
Also, the complete results for the $O(\alpha_s)$ matrix elements of the 
$Q_{1-6}$ operators are not available, although they were recently computed 
in \cite{NLO} in an expansion in powers of $q^2/m_b^2, m_c^2/m_b^2$. These
approximate results are not applicable in the zero recoil region considered
here.

At NNL order the radiative correction from
$\kappa_5$ has to be included as well, which raises the issue of $\gamma_5$
scheme independence of the result. This has been demonstrated explicitly at NLL
order in \cite{BuMu}. In analogy with this result, one expects that the scheme
dependence in $\kappa_5$ will be cancelled by that in the matrix elements of
the four-quark operators
$\sum_{i=1}^6 C_i^{NLL}\langle Q_i\rangle^{NLL}$ and by that in $C_7$. 
In the numerical evaluation one should use for consistency the NDR scheme
for all quantities involved.

The uncertainty in the value of $|V_{ub}|$ extracted from (\ref{1}) coming
from $\Delta(1)$ is dominated by that in ${\cal D}_1(1)$. For 
$|{2 \cal D}_1(1)/(m_b f(1))|\le 0.1 $ one finds from Table I a 3\%
effect in $|V_{ub}|^2$.
A precise computation of ${\cal D}_1(1)$ could help eliminate this source 
of uncertainty.

In the quark model the ratio of form factors ${\cal D}_1(1)/f(1)$ appearing
in (\ref{F1f1}) is always positive. Keeping only a $S$-wave component for the
vector meson wave function, one finds in the static limit for the $b$ quark
\bea
\frac{{\cal D}_1(1)}{f(1)} = \frac{1}{6m_q}\cdot
\frac{\langle \phi_V^\dagger \vec p\,^2 \phi_B\rangle}
{\langle \phi_V^\dagger\phi_B\rangle }\,,
\eea
with $m_q$ the mass of the light quark produced in the weak decay $b\to q$.
The expectation values can be computed explicitly in the ISGW model \cite{ISGW}
with the result
\bea
\frac{{\cal D}_1(1)}{f(1)} = \frac{1}{2m_q}\cdot 
\frac{\beta_B^2\beta_X^2}{\beta_B^2+\beta_X^2} &=& 0.093 \mbox{ GeV}
\quad (B\to \rho)\\
&=& 0.062 \mbox{ GeV}\quad (B\to K^*)\,.\nonumber
\eea
We used here the parameters of the ISGW model $\beta_B= 0.41$ GeV,
$m_u = m_d = 0.33$ GeV, $m_s=0.55$ GeV, $\beta_\rho = 0.31$ GeV,
$\beta_{K^*} = 0.34$ GeV. This amounts to a small positive contribution
of 2-3\% from the last term in (\ref{F1f1}).

\bea\nonumber
\begin{array}{c|rrrrr}
{\cal F}(1)/f(1) & 0.8 & 0.9 & 1.0 & 1.1 & 1.2  \\
\hline
\Delta_{NLL}(1) & -0.123 & -0.137 & -0.151 & -0.164 & -0.178  \\
\Delta_{NNLL}(1) & -0.119 & -0.133 & -0.147 & -0.160 & -0.173  \\
\end{array}
\eea
\begin{quote}
Table 1. The values of the correction factor $\Delta(1)$ appearing in the
formula (\ref{1}) for a few values of the ratio
(\ref{F1f1}) and $\bar\Lambda = 370$ MeV. Two sets of  Wilson coefficients are 
used, corresponding to the
NLL and NNLL approximations, as explained in text. 
\end{quote}

The correction to the ratio of form factors (\ref{F1f1}) can be also extracted from 
the QCD sum rule 
calculation of Ref.~\cite{BaBr}. This gives ${\cal F}(1)/f(1) = 1.17$ for
$B\to \rho$ and 1.18 for $B\to K^*$. A similar QCD sum rule calculation in \cite{ABHH}
quotes the range ${\cal F}(1)/f(1) = 1.15^{+0.13}_{-0.07}$ for $B\to K^*$.
This amounts to a large positive subleading 
contribution of $\sim 30\%$ from the last term in (\ref{F1f1}). 
The large discrepancy in ${\cal D}_1$ with
the dimensional analysis estimate and the quark model result is rather puzzling.
It is not clear if this anomalously large subleading correction is an 
artifact of the sum rule computation or of the interpolation formulas in 
\cite{BaBr,ABHH}. A precise determination of this formfactor is clearly important.

In conclusion, we studied in this Letter the effect of subleading corrections to
a heavy quark symmetry relation relevant for the determination of $|V_{ub}|$ from
exclusive rare and semileptonic $B$ decays. The structure of the
corrections to this symmetry relation is analyzed using the heavy quark 
expansion.
We point out a possible large effect, depending on the value of an 
unknown matrix element of a dimension-4 operator. Lattice computations could
eventually help to eliminate this source of uncertainty in exclusive determinations
of $|V_{ub}|$.

We thank Iain Stewart for comments on the manuscript.
D.~P. is grateful to Andrey Grozin for discussions about 
the results of Ref.~\cite{BrGr}.
This work has been supported by the DOE under Grant No. DOE-FG03-97ER40546.

\end{document}